\documentclass[preprint]{aastex}

\shorttitle{Distant Cluster Hunting}
\shortauthors{Donahue et al.}

\newcommand{\flux}{{\rm erg \, s^{-1} \, cm^{-2}}}

\newcommand{\etal}{{et al. }}
\newcommand{\lum}{{\rm erg \, s^{-1}} }

\newcommand{\ltsim}{{\>\rlap{\raise2pt\hbox{$<$}}\lower3pt\hbox{$\sim$}\>}}
\newcommand{\gtsim}{{\>\rlap{\raise2pt\hbox{$>$}}\lower3pt\hbox{$\sim$}\>}}

\begin{document}
\title{Distant Cluster Hunting: A Comparison Between the Optical
and X-ray Luminosity Functions from an Optical/X-ray Joint Survey}
\author{Megan Donahue\altaffilmark{1}, Jennifer Mack\altaffilmark{1}, 
Caleb Scharf\altaffilmark{1,2}, 
Paul Lee\altaffilmark{1}, Marc Postman\altaffilmark{1}, 
Piero Rosati\altaffilmark{3}, 
Mark Dickinson\altaffilmark{1}, G. Mark Voit\altaffilmark{1}, 
John T. Stocke\altaffilmark{4}} 
\altaffiltext{1}{Space Telescope Science Institute, 3700 San Martin Drive, 
Baltimore, MD 21218, donahue@stsci.edu}
\altaffiltext{2}{Columbia University, Dept. of Astronomy, Mail Code 5246 
Pupin Hall, 550 W. 120th St., New York, NY 10027}
\altaffiltext{3}{European Southern Observatory, Karl-Schwarzschild-Str. 2, 
Garching, D-85748}
\altaffiltext{4}{University of Colorado, CASA, CB 389, Boulder, CO 80309}

\begin{abstract} 
We present a comparison of X-ray and optical luminosities and 
luminosity functions of cluster candidates from a joint optical/X-ray
survey, the ROSAT Optical X-ray Survey (ROXS). 
Completely independent X-ray and optical catalogs of 23 ROSAT 
fields (4.8 square degrees) were created by a matched-filter optical
algorithm and by a wavelet technique in the X-ray.  We directly compare the
results of the optical and X-ray selection techniques. The matched-filter 
technique detected 74\% (26 out of 35) of the most reliable 
cluster candidates in the X-ray-selected 
sample; the remainder could be either constellations of X-ray point
sources or $z>1$ clusters. The matched-filter 
technique identified approximately 3 times the number of
candidates (152 candidates) found in the X-ray survey of the same sky
(57 candidates).  
While the estimated optical  
and X-ray luminosities of clusters of 
galaxies are correlated, the intrinsic scatter in this 
relationship is very large.  We can reproduce the number and distribution
of optical clusters with a model defined by the X-ray luminosity function
and by an $L_x-\Lambda_{cl}$ relation if $H_0=75$~km~s$^{-1}$~Mpc$^{-1}$ 
and if the $L_x-\Lambda_{cl}$ relation is steeper than the expected 
$L_x \propto \Lambda_{cl}^2$. 
On statistical grounds, a
bimodal distribution of X-ray luminous and X-ray faint clusters 
is unnecessary to explain our observations. Followup work is required to confirm whether the clusters without
bright X-ray counterparts are simply X-ray faint for their optical 
luminosity because of their low mass or youth, 
or a distinct population of clusters which do not, for some
reason, have dense intracluster media. 
We suspect that these optical clusters are 
low-mass systems, with correspondingly
low X-ray temperatures and luminosities, or that they are not yet
completely virialized systems.

\end{abstract}


\section{Introduction}

X-ray and optical techniques for detecting clusters of galaxies each
have their merits. Optical techniques have been in use for over four
decades, and images of the optical sky are relatively inexpensive to
obtain. New optical methods such as the matched-filter method 
(Postman et al. 1996, P96 hereafter) 
have allowed for automatic, uniform detection
of optical overdensities of galaxies with magnitude distributions consistent
with those of a typical cluster of galaxies. In the matched-filter
method, fitting galaxy luminosity functions in addition to seeking
overdensities of galaxies minimizes the projection effects
that plagued earlier cluster cataloguing efforts. X-ray selection
has the advantage of directly revealing the hot intracluster medium confined
by the deep gravitational potential of the cluster. Since this emission
is proportional to the gas density squared, and the X-ray sky is sparsely populated compared to the optical, X-ray detections have
higher contrast and less contamination from physically unrelated systems. 
Early X-ray selection methods
optimal for point sources (Gioia et al. 1990b) were biased somewhat towards
selecting clusters with high central surface brightneses, but there
are now several algorithms optimized for detecting extended sources, 
including wavelets and Voronoi-Tesselation Percolation methods (Rosati
et al. 1995; Scharf et al. 1997).

However, X-ray and optical studies of cluster evolution have progressed
along separate paths.
A decade ago, optical and X-ray surveys disagreed in their
assessment of the amount of evolution clusters have experienced. Optical
surveys indicated very little evolution since
$z\sim0.5-1$ (Gunn, Hoessel \& Oke 1986) while X-ray studies suggested
modest (Gioia et al. 1990a) to strong evolution (Edge et al. 1990, later
retracted in Ebeling et al. 1997). 
The most recent X-ray samples of clusters over a range of redshifts out to
$z\sim0.8-1.2$ suggest that the X-ray luminosity function for moderate
luminosity clusters has in fact 
not evolved significantly since $z\sim0.8$ (Borgani et al. 1999; 
Nichol et al. 1999; Rosati et al. 1998; Jones et al. 1998), while the {\em 
most
luminous} (and presumably most massive) systems, 
such as those contained in the EMSS, 
may have evolved somewhat (Gioia et al. 1990a; Henry et al.
1992; Nichol et al. 1997, Vikhlinin et al. 1998, 2000; Rosati et al. 2000;   
Gioia et al. 2001), but the community is not unanimous on
this result (e.g. Jones et al. 1998).  In contrast, recent 
optical surveys for distant clusters continue to find very little
evidence for evolution (Couch et al. 1991; P96).

In an effort to establish the common ground between X-ray and optical studies
of clusters and cluster evolution, we have undertaken a joint X-ray/optical
survey to detect and study clusters of galaxies, called the ROSAT Optical
X-ray Survey, or ROXS. In contrast
to other ROSAT PSPC serendipitous surveys (e.g. Rosati et al. 1995, 
Jones et al. 1998, Romer et al. 2000, Vikhlinin et al. 1998), the ROXS team optically imaged the entire central 30' by 30' of each field in our survey. 
The X-ray selection and optical selection of cluster candidates were then determined 
independently. In this current work, we report results on the relation between the X-ray luminosity ($L_x$) and
$\Lambda_{cl}$, a measure of the cluster's optical luminosity. 
The ROXS has already been used to find a potential intergalactic X-ray
filament (Scharf et al. 2000.)
Paper II (Donahue
et al., 2001) contains the catalogues, data reduction
and observational details. 
For this paper we scale $H_0=75~h_{75}$ km/s/Mpc. We assume
$q_0=0.5$ to ease comparison with earlier results.

\section{Optical Observations and Analysis}

We obtained the optical images  
with the prime focus camera T2KB 
at the Kitt Peak National Observatory 4-meter telescope March, 1996 
and May, 1997. 
The 23 ROXS target fields were selected from a  sample of archival ROSAT PSPC 
observations with exposure times of more than 8,000 seconds and Galactic
latitude of $>20$ degrees. 
Four 900-second, 16' by 16' exposures were obtained through
the I-band filter, to tile the central 30' by 30' of each ROSAT field. 
Our 5$\sigma$ detection limit 
of $I_{AB}=24$ was sufficient to detect cluster galaxies 2 magnitudes fainter
than the typical unevolved first-ranked elliptical at $z=1$ (Postman 
\etal 1998). We constructed galaxy catalogs which were in turn used
to build the optical cluster catalog. 
The matched filter method (P96), tuned
to redshifts in the range $0.2 < z < 1.2$, was used to generate cluster
catalogs in $\Delta z = 0.1$ intervals.
Each cluster candidate is
assigned a central position, radius (corresponding to the area of
detection $= \pi r^2$), an estimated redshift, and a 
a detection confidence (in units of $\sigma$). The algorithm also estimates
an effective optical luminosity 
$\Lambda_{cl}$, which is the equivalent number
of $L^*$ galaxies in the cluster such that the optical cluster 
luminosity in units of solar luminosities is 
$L_{cl}/L_\odot=\Lambda_{cl}L^*=\Lambda_{cl}10^{-0.4(M_\odot - M_*)}$, 
where $M^*=-21.90 + 5 \log h_{75}$ in the I band (P96) and $M_\odot$
is the absolute magnitude of the Sun. A K-correction is applied assuming a standard
elliptical spectrum.  
For reference, a richness class 1 cluster at $z\leq0.7$
has $\Lambda_{cl}=30-65$ in the I-band (P96). 
Additional details regarding field selection, data reduction, and
catalog construction for the galaxies and optical cluster candidates can
be found in Paper II.

We defined 155 optical cluster candidates in 23 30' by 30' fields, of
which 142 satisfy a $\sigma>3$ criterion.
We computed a $4\sigma$ X-ray flux detection threshold within an aperture of
$r=1'$ centered on each optical candidate. 
We also computed the X-ray and optical selection functions 
for the survey (P96, Paper II).

\section{X-ray Cluster Selection and Cross-Identification}

A wavelet-based technique, described by Rosati \etal (1995), was used to
create a catalog of X-ray clusters of galaxy candidates for each field
observed at Kitt Peak. Several of these
fields overlapped with the orginal RDCS (Rosati \etal 1995) sample, and
thus the X-ray cluster candidates in many of the fields already have 
been confirmed and have spectroscopic redshifts. The flux limits are
approximately the same as those of the RDCS ($F_x > 10^{-14} \flux$).
Fifty-seven X-ray candidate clusters and their associated
X-ray parameters were found. 

In order to define the optical/X-ray cross identifications we 
compared locations and contours of X-ray sources to  
contour maps of optical significance
overlaid on the I-band image.  Of the 57 X-ray candidates, 43 
were located within our optical field of view and unobscured by 
scattered light or bright stars. 
Of the 43, 31 were visually 
identified with potential cluster candidates with $\Lambda_{cl} = 25-100$, of which 26 are 
very secure, with centroid separations of $\lesssim 1'$. Of the other 5, 
3 are low-significance optical clusters ($\sigma<3$) which were 
interesting because of their potentially 
high redshift, and two others are more distant affiliations of
sprawling optical systems with a compact X-ray candidate. 
The remaining 12 X-ray cluster candidates can be divided into
4 {\em bona-fide} optically faint candidates, 6 candidates with very
uncertain X-ray fluxes ($F_x/\sigma_{Fx} < 3$), one double source (likely
to be two blended point sources), and one blend with the original ROSAT target.
None of the optically-blank X-ray sources have yet been classified or 
confirmed at other wavelengths. 
Since the spurious fraction typical of the X-ray
surveys is $\sim 10\%$ (Rosati et al. 1998; Vikhlinin et al 1998), 
$\sim6$ of these sources
are likely to be 
spurious or collections of point sources. Some of these candidates may be {\em bonafide},  
albeit optically faint, high-redshift
clusters. Near IR imaging might reveal such clusters.
Ten X-ray candidates have confirmed spectroscopic redshifts; eight of 
these have estimated 
redshifts from the optical data that lie 
within $\Delta z=0.1$ of the spectroscopic redshift (Paper II).
Of the 142 optically selected  
candidates with $\sigma>3$, 27 have X-ray counterparts (one has two 
counterparts).  Up to 29 additional optical candidates have possible X-ray
point-source counterparts within $1'-2'$ 
which we do not use in our correlation analysis.

\section{X-ray Luminosity ($L_x$) and Optical Luminosity ($\Lambda_{cl}$) Relationship}

For a cluster in which mass traces optical light ($M/L_{opt}$ is constant) and  
the gas is in hydrostatic equilibrium ($T \propto M^{2/3}$), 
and $L_x \propto T^3$ (empirical relation see e.g. David et al. 1993), 
we expect the
X-ray luminosity to be related to the optical luminosity 
as $L_x \propto L_{opt}^2$.
The empirical relationship between a cluster's X-ray luminosity and optical
luminosity is not so well defined, in large part due to the difficulties 
inherent in measuring a cluster's optical luminosity and in getting
a homogeneous set of total optical luminosities for a large number of
clusters. Edge \& Stewart (1991) found that the bolometric X-ray
luminosity of a local sample of X-ray selected clusters 
correlated only very roughly with Abell Number and somewhat
better with Bahcall galaxy density (number of bright galaxies
within $0.5h^{-1}$ Mpc; 1977; 1981) 
for the smaller subsample (18) that had Bahcall galaxy densities. 
Arnaud et al.(1992) made a heroic effort to compute cluster optical
luminosities at low redshift from a heterogeneous 
literature. The more systematic matched filter algorithm (P96) 
produces $\Lambda_{cl}$, which is proportional to the estimated optical
luminosity inside $1.5h_{75}^{-1}$~Mpc. 

\begin{figure}
%\plottwo{distant_clusters/fig1a.eps}{distant_clusters/fig1b.eps}
\plottwo{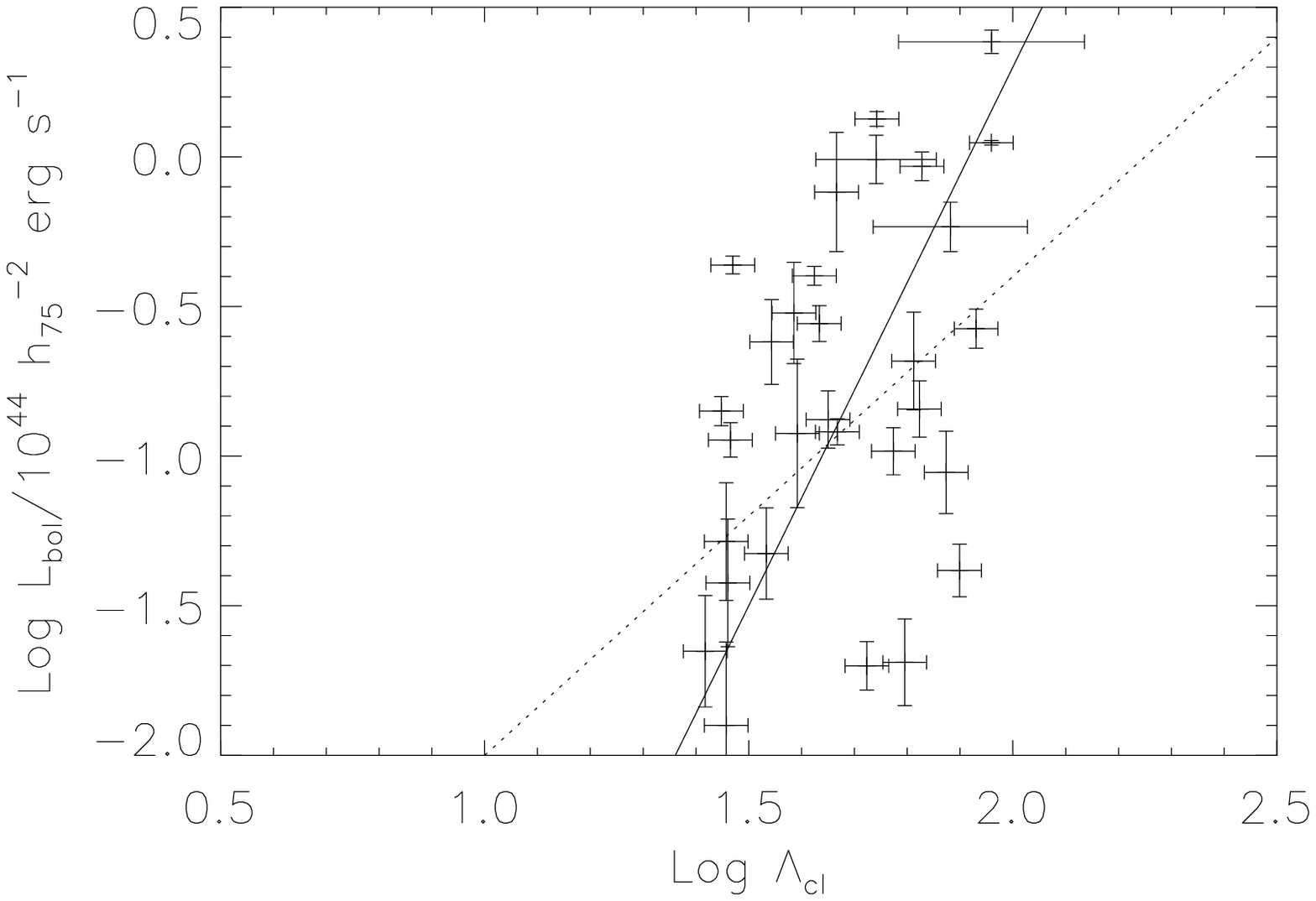}{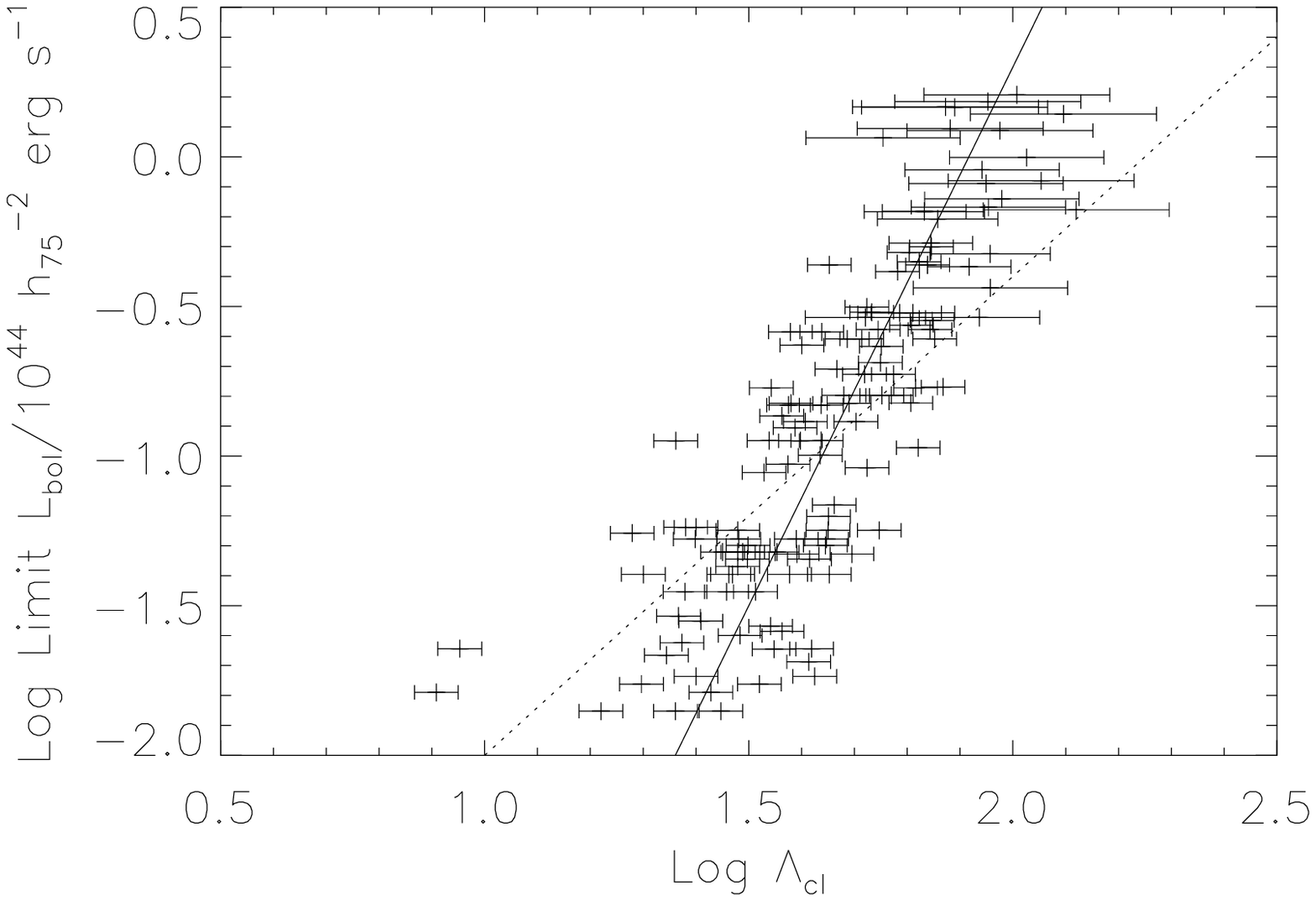}
\caption{We plot bolometric 
$L_x$ (in units of Log $10^{44} h_{75}^{-2} \lum$, 
for $q_0=0.5$) and Log $\Lambda_{cl}$ 
for all jointly detected clusters. The 
dotted line is a best fit to the $L_x-\Lambda_{cl}$ data, 
where $L_x \propto \Lambda_{cl}^{1.6}$. The
solid line represents a best fit obtained in when fit in
in conjunction with the $\Lambda_{cl}$ function and the X-ray
luminosity function, where $L_x \propto \Lambda_{cl}^{3.6}$.
(The fit parameters are only the normalization of the $\Lambda_{cl}$
function and the normalization and slope of the $L_x-\Lambda_{cl}$ relation.) 
On the right hand side, we plot the $4\sigma$ upper limits 
to $L_x$ and the uncertainty on $\Lambda_{cl}$, with the same
line key as on the right.  
The moderate bolometric corrections for fluxes in the 
observed ROSAT bandpass were based on the estimated $T_x$
for the cross-identified candidates and a maximum $T_x$ for the 
upper limits, iteratively computed from $L_x$ and the $L_x-T_x$ relation from
Markevitch (1998). \label{lx_lambda}}
\end{figure}

We plot the the relationship between bolometric 
$L_x$ (in units of $10^{44} h_{75}^{-2} \lum$) and $\Lambda_{cl}$ for
our sample of clusters of galaxies in Figure~\ref{lx_lambda}, including
the $4\sigma$ 
upper limits for clusters for which there was no X-ray counterpart. 
We tested the relationship between $L_x$, $\Lambda_{cl}$ and 
estimated $z$, including the upper limits for $L_x$, with a 
statistical procedure by Akritas \& Siebert (1996) and 
Kembhavi, Feigelson, \& Singh (1986) to test for
partial correlation in the presence of censored data. Our aim was 
to see if $L_x$ and $\Lambda_{cl}$ were indeed correlated beyond the 
effect of $z$ estimates on flux-limited surveys.
We found that while the correlation between $L_x$ and $z$ is 
fairly strong, with a Kendall $\tau$-coefficient of 0.6, 
the correlation between $\Lambda_{cl}$ and $z$
is not nearly as strong, $\tau=0.037$, and the partial correlation
of all 3 quantities is significant at the 95\% level (Partial
$\tau=0.057\pm0.019$). The correlation between $L_x$ and $\Lambda_{cl}$ is
$\tau=0.0677\pm0.019$ (the same $\sigma_\tau$ for all $\tau$,
Kembhavi et al. 1986), and
thus $L_x$ and $\Lambda_{cl}$ are correlated at the $3\sigma$ level.

We roughly fit a correlation of
$\log L_{44} = (-3.6 \pm 0.7) + (1.6 \pm 0.4) \log \Lambda_{cl}- 2\log h_{75}$,
where $L_{44}=L_{x,bol}/10^{44} \lum$, to the 
cross-identified X-ray/optical clusters using a method of
bivariated correlated errors and scatter (Akritas \& Bershady 1996) 
with bootstrap estimate of the variance, including intrinsic
scatter. This fit is by no means a 
unique fit to the data, but it suits our purposes for the next 
discussion. The fit and scatter are also consistent with the upper limits, which 
are well mixed with the detections.  
The intrinsic scatter in the relation and the fit is significant.

\section{The $\Lambda_{cl}$ Function - X-ray Luminosity Function Comparison
\label{lf}}

\begin{figure}
%\plotone{lambda/dndlamh75dndlx.eps}
\plotone{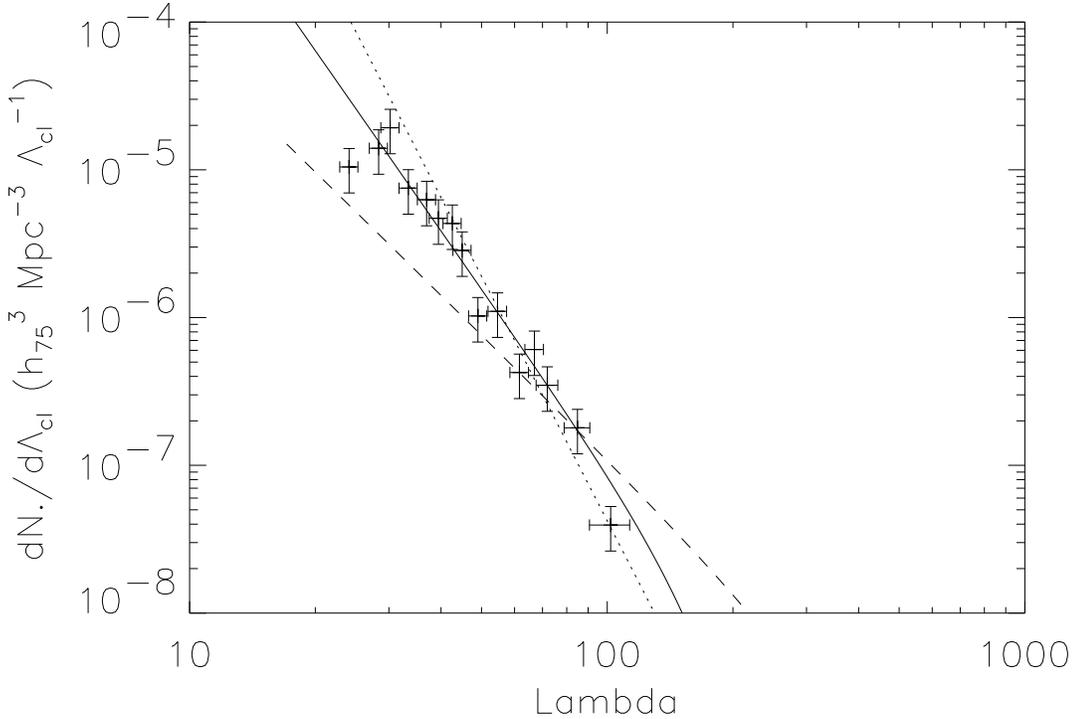}
\caption[]{Binned differential $\Lambda_{cl}$ function for the ROXS  
clusters, with no correction for spurious clusters. The error bars 
represent the weighted observational uncertainty of the mean $\Lambda_{cl}$
in that bin of $\sim9$ clusters each. The dotted line
represents the best fit power-law obtained
through a maximum likelihood method for the unbinned sample. 
The solid line is the best fit $\Lambda_{cl}$
function as derived from the X-ray luminosity function and the
simultaneously fit $\Lambda_{cl}-L_x$ relation with $L_x \propto 
\Lambda_{cl}^{3.6}$. If $L_x \propto \Lambda_{cl}^{1.6}$, then the
inferred $\Lambda_{cl}$ function is represented by the 
dashed line. Including moderate scatter changes the normalization
but not the slope of these relations. For
this plot, we have used $q_0=0.5$ and $H_0=75 h_{75}$ km s$^{-1}$ Mpc$^{-1}$.  
\label{diff_fn}}
\end{figure}

We can compute the observed ``$\Lambda_{cl}$-Function'' for these data 
from the estimates of $\Lambda_{cl}$, the selection function (a function of $\Lambda_{cl}$ and redshift),  
and the effective sky coverage of the
survey (4.84 square degrees). 
The maximum search volume as a function of $\Lambda_{cl}$ is 
estimated for each cluster. (For details, see Paper II.) For
a given $\Lambda_{cl}$, the $N(>\Lambda_{cl})$ is the sum over all $1/V_{max}(\Lambda_{cl})$ 
for which $\Lambda_{cl} > \Lambda$. We plot  
the differential function in Figure~\ref{diff_fn}, binned such that each
bin but the highest contains 9 clusters. This function
is consistent with that derived from the much larger sample in the
Deeprange survey (Postman et al, in preparation.)

% /data/atalanta1/megan/papers/rox/lambda/max_like_lambda.pro
% fits and plots N(>Lambda)

We obtain a raw estimate of $N$ 
at $\Lambda_{cl}\gtsim60$ of $\sim 1.3\times10^{-5} h_{75}^3$ Mpc$^{-3}$,
and $\Lambda_{cl}\gtsim80$ of $\sim 4 \times 10^{-6} h_{75}^3$ Mpc$^{-3}$.
The estimated value is consistent with the value for the 
Postman \etal (1996) survey at the same $\Lambda_{cl}$,  
($\sim4^{+3}_{-2}\times 10^{-6} h_{75}^{3} ~\rm{Mpc}^{-3}$).  The 
best fit differential $\Lambda_{cl}$ function ($H_0=75$, $q_0=0.5$)
is of the form:
\begin{equation}
\frac{dn}{d\Lambda_{cl}} = n_0 ( \frac{\Lambda_{cl}}{40})^{-\alpha}
\end{equation}
where $\alpha = 5.3 \pm 0.5$ and $n_0=6^{+3}_{-1} \times 10^{-6} h_{75}^3
\, {\rm Mpc}^{-3} (\Lambda_{cl})^{-1}$. The uncertainties here do not 
take into account the uncertainties on the survey volume, 
$\Lambda_{cl}$, or spurious fraction. Each  ($\Lambda_{cl}$, $z$)  
data point is weighted equally in the fit and the influence of the 
spurious fraction 
is minimized by fitting only those 134 cluster candidates 
with $\Lambda_{cl}\ge 20$ and $\sigma \ge 3$.

Are the optical cluster candidates without X-ray counterparts spurious,
real but X-ray faint clusters, or the high $\Lambda_{cl}$-tail of 
the more numerous low-mass, low-$L_x$ X-ray clusters? We know already from
spectroscopic followup of similar matched filter surveys that these
clusters cannot all be spurious 
(Holden, et al. 1999; Adami et al 2000; Postman, Lubin \& Oke 1998). 
We compare the $\Lambda_{cl}$ 
function to the X-ray luminosity function (XLF) 
for $0.1-2.4$ keV
of the X-ray clusters of galaxies from the Brightest
Cluster Survey (Ebeling \etal 1997) 
in order to answer this question.

A chain-rule conversion
of the XLF to a $\Lambda_{cl}$ function using
the $L_x-\Lambda_{cl}$ relation fit in \S3 produces a $\Lambda_{cl}$ 
function significantly 
flatter than
the observed $\Lambda_{cl}$ function (Figure~\ref{diff_fn}). If we 
simultaneously fit {\em both} the $\Lambda_{cl}$
distribution function and the $L_x-\Lambda_{cl}$ relation, we obtain
a steeper relation where $\log (L_{bol,44} h_{75}^{2}) =  (- 6.9\pm1.1)
+ (3.6\pm0.8) \log \Lambda_{cl}$, where the best-fit values of the
slope and normalization are highly correlated. 
This relation, when used to convert
the XLF to the $\Lambda_{cl}$ function, results in a steeper 
$\Lambda_{cl}$ function.
We have also investigated the
effect of a Gaussian distribution in $\Lambda_{cl}$ with 
respect to $L_x$ for the inferred $\Lambda_{cl}$ function. The qualitative
effect of scatter is to change the normalization but not the slope of
the $\Lambda_{cl}$ function. 

The optical matched filter method detected many more cluster
candidates than did the X-ray method, at threshholds optimal 
for maximizing the number of cross-correlated 
sources (Paper II). Our analysis demonstrates 
that at least one of the following three assumptions is false: 
(a) $L_x \propto
\Lambda_{cl}^2$, (b) that we have accurately estimated the $\Lambda_{cl}$
distribution function from the present dataset, 
and/or (c) that the X-ray luminosity function is 
accurately known. Of these three, (c) is most likely to be 
true, as several groups have obtained compatible results. The
assumption (b) can be confirmed by an independent matched-filter
survey (Deeprange, Postman et al. 1998). So (a) is the most likely weak link.

Our results are consistent with the hypothesis that all of
the optical cluster candidates with $\Lambda_{cl}>30$ are true clusters, 
as long as the true $L_x-\Lambda_{cl}$ relation is rather
steep ($L_x \propto \Lambda_{cl}^{3.8\pm0.8}$).  
We do not need to assume that the 
optical candidates not detected in X-rays are 
X-ray faint yet massive clusters to explain our results here. 
However, if the $L_x-\Lambda_{cl}$ relation is indeed steeper than $\beta=2$ 
and $\Lambda_{cl}$ is directly proportional to $L_{opt}$, this result
implies that the mass to light ratio of clusters continues to increase
with the mass of the cluster, at least into the moderate mass range 
explored in our sample. We note that our survey is too small to 
include the most massive clusters, and our estimates of $L_x$ and
$L_{opt}$ are confined to the central few arcminutes of each
candidate. If $L_x \propto M^2$ at these mass scales, 
$M/L_{opt} \propto M^{1-2/\beta}$ 
for the range of cluster masses in our survey ($\sim 10^{14}-10^{15} 
h_{75}^{-1} M_\odot$). This weak dependence of $M/L$ on mass is not ruled
out by existing data (e.g. Hradecky et al. 2000.)

Followup with weak lensing measurements, near-IR, and X-ray 
observations are essential.  
If the clusters are low-mass, the clusters with high 
$\Lambda_{cl}$ but low $L_x$ will have commensurately low X-ray temperatures.
ROXS identifies optically rich examples
of cluster candidates without extended X-ray counterparts for the 
assessment of the possible existence of optically
bright but X-ray faint clusters of galaxies.

\section{Summary}

We have directly compared the results of cluster hunting with two 
competing methods: optical selection in the I-band, 
using the matched filter technique, and
X-ray selection of extended sources. We have found that both methods 
reliably detect most of the richest, and presumably most massive systems, 
but the optical matched filter technique, because of the scatter and 
the steepness of the relation between X-ray and optical luminosities, 
produces more cluster candidates at our sensitivities.  
We present the first $\Lambda_{cl}$ function for 
optically-selected clusters of galaxies. 
The $\Lambda_{cl}$ function is consistent with
the X-ray luminosity function for clusters of galaxies if the 
intrinsic, global $\Lambda_{cl}-L_x$ relation is consistent  
with that of the cross-identified candidates.
In particular, we can explain our observations without appealing to 
an X-ray faint population of massive clusters of galaxies. 

Predictions 
for an observational test using deep XMM-Newton/EPIC exposures 
based on our result are: 
(1) X-ray observations of the undetected, high-$\Lambda_{cl}$
systems should detect X-rays from most of the systems with 
observations of only moderately increased sensitivity over the ROSAT observations, 
and the detected ICM will not be hot ($T_x\lesssim4$ keV). 
(2) If $L_x \propto \Lambda_{cl}^{3-4}$, the median predicted
bolometric X-ray luminosity for the entire undetected sample 
is $\sim 10^{43}h_{75}^{-2}$ erg s$^{-1}$, so that deep
XMM-Newton EPIC exposures should detect most of this population. On the
other hand, significant numbers of non-detections with XMM will suggest
that many of these clusters are spurious or are significantly less
massive than their estimated richnesses may suggest or that there 
is an X-ray faint population at these mass scales.
(3) The X-ray 
candidates which are not detected by the matched filter algorithm 
in the optical images will be revealed
to be either false clusters of galaxies (constellations of AGN), 
moderately distant groups, or distant clusters of galaxies ($z>1-1.2$).

\end{document}